\documentclass[aps,pre,amsmath,amsfonts,floatfix,superscriptaddress,twocolumn]{revtex4}

\usepackage{graphicx}

\font\tenmib=cmmib10
\font\sevenmib=cmmib10 scaled 800

 2

\font\sc=cmcsc10

\font\ottorm=cmr8
\font\msytw=msbm9 scaled\magstep1

\font\indbf=cmbx10 scaled\magstep2

\font\ottorm=cmr8\font\ottoi=cmmi8\font\ottosy=cmsy8%
\font\ottobf=cmbx8\font\ottott=cmtt8%
\font\ottocss=cmcsc8%
\font\ottosl=cmsl8\font\ottoit=cmti8%
\font\sixrm=cmr6\font\sixbf=cmbx6\font\sixi=cmmi6\font\sixsy=cmsy6%
\font\fiverm=cmr5\font\fivesy=cmsy5
\font\fivei=cmmi5
\font\fivebf=cmbx5%

\def\ottopunti{\def\rm{\fam0\ottorm}%
\textfont0=\ottorm\scriptfont0=\sixrm\scriptscriptfont0=\fiverm%
\textfont1=\ottoi\scriptfont1=\sixi\scriptscriptfont1=\fivei%
\textfont2=\ottosy\scriptfont2=\sixsy\scriptscriptfont2=\fivesy%
\textfont3=\tenex\scriptfont3=\tenex\scriptscriptfont3=\tenex%
\textfont4=\ottocss\scriptfont4=\sc\scriptscriptfont4=\sc%
\textfont5=\tenmib\scriptfont5=\sevenmib\scriptscriptfont5=\fivei
\textfont\itfam=\ottoit\def\it{\fam\itfam\ottoit}%
\textfont\slfam=\ottosl\def\sl{\fam\slfam\ottosl}%
\textfont\ttfam=\ottott\def\tt{\fam\ttfam\ottott}%
\textfont\bffam=\ottobf\scriptfont\bffam=\sixbf%
\scriptscriptfont\bffam=\fivebf\def\bf{\fam\bffam\ottobf}%
\setbox\strutbox=\hbox{\vrule height7pt depth2pt width0pt}%
\normalbaselineskip=9pt\let\sc=\sixrm\normalbaselines\rm}
\let\a=\alpha \let\b=\beta    \let\d=\delta 
     \let\th=\theta  
\let\m=\mu             \let\p=\pi    \let\r=\rho
    \let\f=\varphi 
   
\let\G=\Gamma \let\D=\Delta  \let\Th=\Theta 
    \let\Si=\Sigma     \let\Ps=\Psi
\let\O=\Omega 

\def\\{\hfill\break} \let\==\equiv

\let\io=\infty 

\let\0=\noindent

\def\ie{{i.e. }}

\def\tende#1{\,\vtop{\ialign{##\crcr\rightarrowfill\crcr
 \noalign{\kern-1pt\nointerlineskip} \hskip3.pt${\scriptstyle
 #1}$\hskip3.pt\crcr}}\,}
\def\circage{\lower2pt\hbox{$\,\buildrel > \over {\scriptstyle \sim}\,$}}
\def\otto{\,{\kern-1.truept\leftarrow\kern-5.truept\to\kern-1.truept}\,}

\def\NN{{\cal N}}

\def\T#1{{#1_{\kern-3pt\lower7pt\hbox{$\widetilde{}$}}\kern3pt}}
\def\VVV#1{{\underline #1}_{\kern-3pt
\lower7pt\hbox{$\widetilde{}$}}\kern3pt\,}
\def\W#1{#1_{\kern-3pt\lower7.5pt\hbox{$\widetilde{}$}}\kern2pt\,}

\def\indica{\leaders \hbox to 0.5cm{\hss.\hss}\hfill}
\def\guida{\leaders\hbox to 1em{\hss.\hss}\hfill}

\def\ul{\underline}

\def\erf{\text{erf}}

\mathchardef\aa   = "050B
\mathchardef\bb   = "050C
\mathchardef\ggg  = "050D
\mathchardef\xxx  = "0518
\mathchardef\zzzzz= "0510
\mathchardef\oo   = "0521
\mathchardef\lll  = "0515
\mathchardef\mm   = "0516
\mathchardef\Dp   = "0540
\mathchardef\H    = "0548
\mathchardef\FFF  = "0546
\mathchardef\ppp  = "0570
\mathchardef\nn   = "0517
\mathchardef\ff   = "0527
\mathchardef\pps  = "0520
\mathchardef\FFF  = "0508
\mathchardef\nnnnn= "056E

\def\to{\rightarrow}

\def\qed{\raise1pt\hbox{\vrule height5pt width5pt depth0pt}}

\def\indic{\hbox{\raise-2pt \hbox{\indbf 1}}}

\def\RRR{\hbox{\msytw R}}

\def\ul#1{{\underline#1}}

\def\V0{{\bf 0}}

\def\V#1{{\underline#1}}

\newcommand{\beq}{\begin{equation}}
\newcommand{\eeq}{\end{equation}}
\newcommand{\wh}{\widehat}

\newcommand{\Tr}{\text{Tr}}

\begin{document}

\title{The ideal glass transition of Hard Spheres}

\author{Giorgio Parisi\footnote{giorgio.parisi@roma1.infn.it}
}

\affiliation{Dipartimento di Fisica, Universit\`a di Roma ``La Sapienza'', 
P.le A. Moro 2, 00185 Roma, Italy}

\affiliation{INFM -- CRS SMC, INFN, Universit\`a di Roma ``La Sapienza'', 
P.le A. Moro 2, 00185 Roma, Italy}

\author{Francesco Zamponi\footnote{francesco.zamponi@phys.uniroma1.it}
}

\affiliation{Dipartimento di Fisica, Universit\`a di Roma ``La Sapienza'', 
P.le A. Moro 2, 00185 Roma, Italy} 

\affiliation{INFM -- CRS Soft, Universit\`a di Roma ``La Sapienza'', 
P.le A. Moro 2, 00185 Roma, Italy}

\date{\today}

\begin{abstract} 
We use the replica method to study the ideal glass transition
of a liquid of identical Hard Spheres. We obtain estimates of the 
configurational entropy in the liquid phase, of the
Kauzmann packing fraction $\f_K$, in the range $0.58 \div 0.62$,
and of the random close packing density $\f_c$, in the range
$0.64 \div 0.67$, depending on the approximation we use for the equation 
of state of the liquid.
We also compute the pair correlation function in the glassy
states ({\it i.e.}, dense amorphous packings) and we find
that the mean coordination number at $\f_c$ is equal to $6$.
All these results compare well with numerical simulations and
with other existing theories.
\end{abstract} 

\maketitle

\section{Introduction}

The question whether a liquid of identical Hard Spheres undergoes
a glass transition upon densification is still open
\cite{RT96,RLSB98,Sp98,TdCFNC04}.
If crystallization is avoided, one can access 
the metastable region of the phase diagram above the freezing
packing fraction $\f_f=0.494$, where $\f=\frac{N \p D^3}{6V}$, 
$D$ is the Hard Sphere diameter, $N$ is the number of particles
and $V$ is the volume of the container.
In this region the dynamics of the liquid becomes slower and slower
on increasing the density. The particles
are ``caged'' by their neighbors, and the dynamics separates into
a fast rattling inside the cage and slow rearrangements of the
cages. The typical time scale of these rearrangements increase
very fast around $\f_g \sim 0.56$ and many authors reported the
observation of a glass transition at these values of 
density~\cite{GS91,vMU93}.

If the radius of the cages is sufficiently small and if the
typical time scale of cage rearrangements is sufficiently
large, the system vibrates around configurations
that are stable for a very large time and can be threated as
metastable states. It is then natural to separate the total
entropy of the liquid in a ``vibrational'' contribution, that
accounts for the entropy related to the rattling of the particles
around the metastable structure, and a ``configurational'' entropy
that is the number of metastable states accessible to the liquid
at the considered value of density \cite{SW84, DeB96}.
For many simple potentials such as the Lennard--Jones~\cite{CMPV99,SKT99} 
and for more realistic systems as well \cite{Ka48,An95} 
the extrapolation of the measured
configurational entropy at higher density (or lower temperature) 
indicates that there exists a density, 
called {\it Kauzmann density} $\f_K$, where
the configurational entropy vanishes. The system freezes in the
lowest free-energy states and no more rearrangements of the
structure are possible. This transition is commonly called 
{\it ideal glass transition} or {\it Kauzmann transition}
\cite{DeB96,CMPV99,SKT99,Ka48,An95,MP99}. Note that the Kauzmann density
is expected to be larger than the experimental glass transition density,
as at $\f_K$ the relaxation time is expected to diverge so that
the system freezes in a metastable state, on the experimental time
scale, for a density $\f_g$ smaller than $\f_K$.
The density $\f_g$ where the real glass transition happens
(weakly) depends on the experimentally accessible time scale.
Few estimates of the configurational entropy for Hard Spheres
are currently available \cite{Sp98,CFP98,Luca05} and indicate
a value of $\f_K$ in the range $0.58 \div 0.62$.

A related problem is the study of {\it dense amorphous packings}
of Hard Spheres. Dense amorphous packings are relevant in 
the study of colloidal suspensions, granular matter, powders,
etc. and have been widely studied in the 
literature~\cite{Be83,SK69,Fi70,Be72,Ma74,Po79,Al98,SEGHL02}.
The amorphous metastable configurations described above provide
examples of such packings: when the system freezes in one of these
states, if one is still able to increase the density in order to
reduce the size of the cages to zero (for example by shaking the
container \cite{SK69,Fi70} or making use of suitable computer 
algorithms \cite{Be72,Ma74,SEGHL02}), a {\it random close packed}
state is reached.
The problem of which is the maximum value of density $\f_c$ that
can be reached applying this kind of procedures has been tackled using
a lot of different techniques, usually finding values of $\f_c$ in the
range $0.62 \div 0.67$.
Another interesting problem is to estimate the mean
coordination number $z$, {\it i.e.} the mean number of contacts between
a sphere and its neighbors, in the random close packed states.
Many studies addressed this question usually finding values of 
$z \sim 6$.

Recently, the replica method \cite{MPV87,Mo95,MP99}
has been successfully applied to the
study of the ideal glass transition in simple liquids as the 
Lennard--Jones liquid. 
Reliable estimates of the configurational
entropy, of the Kauzmann temperature and of the thermodynamic 
properties of the glass have been obtained from first principles
in this way~\cite{MP99,MP99b,CMPV99,MP00}. 
However, for technical reasons
this approach could not be extended straightforwardly to the
case of Hard Spheres; indeed at some stage is was assumed that the vibrations around the
equilibrium positions were harmonic in a first approximation. This approximation is not bad for soft 
potentials, but it clearly makes no sense for hard spheres.
A related but different approach 
was used in \cite{CFP98}, obtaining a reasonable estimate
of the Kauzmann density $\f_K \sim 0.62$; however, the estimate
of the configurational entropy was wrong by two orders of 
magnitude and the thermodynamic properties of the glass could
not be computed within this approach.

The aim of this work is to adapt the replica method of 
\cite{MP99} to the case of the Hard Sphere liquid, and 
in general of potentials such that the pair distribution
function $g(r)$ shows discontinuities.
This allows us to compute from first principles the 
configurational entropy of the liquid as well as 
the thermodynamic properties of the glass and the random 
close packing density.
We find a very good estimate of the configurational 
entropy that agrees well with recent numerical simulations
\cite{Sp98,Luca05}, a Kauzmann density in the range $0.58 \div 0.62$
(depending on the equation of state we use to describe
the liquid state), and a random close packing density in the
range $0.64 \div 0.67$. Moreover, we find that the mean 
coordination number in the amorphous packed states is $z=6$
irrespective of the equation of state we use for the liquid,
in very good agreement with the result of numerical 
simulations~\cite{Be72,Ma74,SEGHL02}.

The structure of the paper is the following:
in section~\ref{sec:replica} we outline the replica method
of~\cite{MP99}; in section~\ref{sec:smallcageexp} we show
how it can be adapted to the case of Hard Spheres;
in section~\ref{sec:freeenergy} we resume the main formulae
from which we derive our results;
in section~\ref{sec:results} we present our main results 
about the configurational entropy of the liquid and the
thermodynamic properties of the glass;
in section~\ref{sec:correlations} we discuss the behavior of
the correlation functions in the glass phase;
finally, in section~\ref{sec:discussion} we compare our results
with previous works.

\section{The replica approach to the structural glass transition}
\label{sec:replica}

The replica method was successfully adapted to the study
of the glass transition of simple liquids in a series of
recent papers \cite{Mo95,MP99,MP99b,MP00,CMPV99}. 
The strategy as well as the physics
beyond it have been described in detail in~\cite{MP99}:
in this section we will only review the main steps of
this approach in order to establish some notations.

\subsection{The molecular liquid}

Let us consider here a system at fixed density as in~\cite{MP99}.
The discussion is trivially extended to the case of interest here
where the density is the control parameter.

Close to the glass transition the phase space is disconnected in
an exponential number of states. The number of states of free energy
$f$ is called $\NN(f) = \exp N \Si(f)$. The complexity $\Si(f)$ is
a concave function of $f$ and vanishes at some value $f_{min}$.
One can write the 
partition function $Z$ in the following way:
\beq
\label{Zm1}
\begin{split}
Z &= e^{-\b N F(T)} \sim \sum_\a e^{-\b N f_\a} \\
&= \int_{f_{min}}^{f_{max}}df \, e^{N [\Si(f)-\b f]}
\sim  e^{N [\Si(f^*)-\b f^*]} \ ,
\end{split}
\eeq
where $f^*$ is such that $\b \Phi(f)=\b f - \Si(f)$ is minimum.
The ideal glass transition is met at the temperature $T_K$ such that $f^*(T_K) = f_{min}$ 
and $\Si(f^*)=0$.

The basic idea of the replica approach \cite{Mo95,MP99}
is to consider $m$ copies of the original system, constrained to be in the same state by
a small attractive coupling. The partition function of the replicated system is then
\beq
\label{Zm}
\begin{split}
Z_m &= e^{-\b N \Phi(m,T)} \sim \sum_\a e^{-\b N m f_\a} \\
&= \int_{f_{min}}^{f_{max}}df \, 
e^{N [\Si(f)-\b m f]}
\sim  e^{N [\Si(f^*)-\b m f^*]} \ ,
\end{split}\eeq
where now $f^*(m,T)$ is such that $\b \Phi(m,f)=\b m f - \Si(f)$ is minimum.
If $m$ is allowed to assume real values, the complexity can be estimated from the knowledge
of the function $\b \Phi(m,T)=\b m f^*(m,T) - \Si(f^*(m,T))$. 
Indeed, it is easy to show that
\beq
\label{mcomplexity}
\begin{split}
\b f^*(m,T) &= \frac{\partial \, \b \Phi(m,T)}{\partial m} \ , \\
\Si(m,T) &= \Si(f^*(m,T)) = m^2 \frac{\partial \,[ m^{-1} \b \Phi(m,T)]}{\partial m} \\&= 
m \b f^*(m,T) - \b \Phi(m,T) \ .
\end{split}
\eeq
The function $\Si(f)$ can be reconstructed from the parametric plot of $f^*(m,T)$ and $\Si(m,T)$.

Moreover, at fixed $m < 1$, the glass transition is shifted towards
lower values of the temperature. Indeed, for any value of the temperature $T$ below $T_K$
it exists a value $m^*(T) < 1$ such that for $m < m^*$ the system is in the liquid phase.
The free energy for $T<T_K$ and $m < m^*(T)$ can be computed by analytic continuation of 
the free energy of the high temperature liquid. As the free energy is always continuous
and it is {\it independent} of $m< m^*(T)$ in the glass phase (being simply the value $f_{min}(T)$
such that $\Si(f_{min})=0$), one can compute the free energy of the glass below $T_K$
simply as $F_{glass}(T)=\Phi(m^*(T),T)/m^*(T)$.

The $m$ copies are assumed to be in the same state.  This means that each atom of a given replica is
close to an atom of each of the other $m-1$ replicas, {\it i.e.}, the liquid is made of {\it
molecules} of $m$ atoms, each belonging to a different replica of the original system.  In other
words the atoms of different replicas stay in the same cage.  The replica method allow us to define
and compute the properties of the cages in a purely equilibrium framework, in spite of the fact that
the cages have been defined originally in a dynamic framework.  The problem is then to compute the
free energy of a molecular liquid where each molecule is made of $m$ atoms.  The $m$ atoms are kept
close one to each other by a small inter-replica coupling that is switched off at the end of the
calculation, while each atom interacts with all the other atoms of the same replica via the original
pair potential.  This problem can be tackled by mean of the HNC integral equations~\cite{Hansen}.

\subsection{HNC free energy}

The traditional HNC approximation can be naturally extended to the case where particles have internal degrees of
freedom and also to the replica approach where we have molecules composed by $m$ atoms.

We will denote by $x=\{\ul x_1,\cdots,\ul x_m\}$, $\ul x_a \in \RRR^d$ the coordinate of a molecule
in dimension $d$.  The single-molecule density is
\beq
\r(x) = \langle \sum_{i=1}^N \prod_{a=1}^m \delta (\ul x_{ia} - \ul x_a) \rangle \ ,
\eeq
and the pair correlation is
\beq
\r(x) g(x,y) \r(y) = \langle \sum_{i,j}^{1,N} \prod_{a=1}^m \delta (\ul x_{ia} - \ul x_a)
\prod_{b=1}^m \delta (\ul x_{jb} - \ul y_b)  \rangle \ .
\eeq
We define also $h(x,y)=g(x,y)-1$. The interaction potential between two molecules
is $v(x,y)= \sum_a v(|\ul x_a-\ul y_a|)$.

The HNC free energy is given by~\cite{Hansen,MP99}
\beq
\label{HNCfree}
\begin{split}
&\b \Ps[\r(x),g(x,y)] = \frac{1}{2} \int dx dy \, \r(x) \r(y) 
\big[ g(x,y) \log g(x,y) \\
&\hskip20pt - g(x,y) + 1 + \b v(x,y) g(x,y) \big] \\
&+ \int dx \r(x) \big[ \log \r(x) -1 \big]
+ \frac{1}{2} \sum_{n\geq 3} \frac{(-1)^n}{n} \Tr [ h\r]^n \ ,
\end{split}
\eeq
where
\beq\begin{split}
\Tr [ h\r]^n &= \int dx_1 \cdots dx_n h(x_1,x_2) \r(x_2) h(x_2,x_3) \r(x_3) \\
&\cdots h(x_{n-1},x_n) \r(x_n) h(x_n,x_1) \r(x_1) \ .
\end{split}\eeq
For Hard Spheres the potential term vanishes, 
$\int dx dy \, \r(x)\r(y) g(x,y)v(x,y) \equiv 0$, so the
reduced free energy $\b \Psi$ will not depend on the 
temperature in all the following equations.  Similarly, all the free energy functions that we will
consider below do not depend on the temperature once multiplied by $\b$.  In principle we could
stick to $\beta=1$ and slightly simplify the formulae.  We have preferred to keep explicitly
$\beta$, in order to conform to the standard notation for soft spheres (or for hard spheres with an
extra potential).\\

Differentiation w.r.t $g(x,y)$ leads to the HNC equation:
\beq
\log g(x,y) + \b v(x,y) = h(x,y)-c(x,y) \ ,
\eeq
having defined $c(x,y)$ from
\beq
h(x,y)=c(x,y)+\int dz \, c(x,z)\r(z)h(z,y) \ .
\eeq
The free energy (per particle) of the system is given by
\beq\begin{split}
&\phi(m,T)= \frac{1}{Nm} \min_{\r(x), g(x,y)} \Psi[\r(x),g(x,y)] \ , \\
&\Phi(m,T) = m \phi(m,T) \ ,
\end{split}\eeq
and once the latter is known one can get the free energy of the states and the complexity
using Eq.s~(\ref{mcomplexity}).

\subsection{Single molecule density}

The solution of the previous equations for generic $m$ is a very complex problem (it is already
rather difficult for $m=2$). Some kind of {\it ansatz} is needed to simplify the computation, that may
become terribly complicated.

The single molecule density encodes the information about the inter-replica coupling
that keeps all the replicas in the same state. We assume that this arbitrarily small coupling
has already been switched off, with the main effect of building molecules of $m$
atoms vibrating around the center of mass $\ul X \in \RRR^d$ of the molecule with a certain 
``cage radius'' $A$. The simplest {\it ansatz} for $\r(x)$ is then~\cite{MP99}
\beq
\label{rrho}
\r(x) = \wh\r \int d \ul X \prod_a \r(\ul x_a-\ul X) \ , \hspace{10pt} \int d \ul u \, \r(\ul u)=1 \ ,
\eeq
with
\beq
\label{rho}
\r(\ul u)=\frac{e^{-\frac{u^2}{2A}}}{(\sqrt{2\p A})^d} \ ,
\eeq
and $\wh\r=V^{-1} \int dx \, \r(x)$ the number density of molecules.
With this choice it is easy to show that
\beq\begin{split}
\frac{1}{N} \int & dx \, \r(x) \big[ \log \r(x) -1 \big] =
\log \wh\r -1 +\\
&\frac{d}{2} (1-m) \log ( 2\p A ) - \frac{d}{2} \log m + \frac{d}{2}(1-m)
\end{split}\eeq

\subsection{Pair correlation}

As the information about the inter-replica coupling is already encoded
in $\r(x)$, we make the {\it ansatz} for $g(x,y)$:
\beq
\label{gprod}
g(x,y) = \prod_a g(| \ul x_a-\ul y_a |) \ ,
\eeq
where $g(r)$ is rotationally invariant because so is the interaction
potential.
We also define $G(r) \equiv [g(r)]^m$.
Using the {\it ansatz} above, it is easy to rewrite the free
energy (\ref{HNCfree}) as follows:
\beq
\label{HNCfree2}
\begin{split}
\b \Ps &= \frac{\wh\r N}{2} \int d \ul r \, 
\big\{m [F_0(r)]^{m-1} F_1(r) - [F_0(r)]^m \\&+ 1
+ m  [F_0(r)]^{m-1} F_v(r) \big\} \\
&+ \int dx \, \r(x) \big[ \log \r(x) -1 \big]
+ \frac{1}{2} \sum_{n\geq 3} \frac{(-1)^n}{n} \Tr [ h\r]^n \ ,
\end{split}
\eeq
where
\beq
\label{Fp}
\begin{split}
&F_p(|\ul r|) = \int d\ul u d\ul v \, \r(\ul u)\r(\ul v) \, 
g(|\ul r+\ul u-\ul v|) [\log g(|\ul r+\ul u-\ul v|)]^p \\
&F_v(|\ul r|) = \int d\ul u d\ul v \, \r(\ul u)\r(\ul v) \, 
g(|\ul r+\ul u-\ul v|) \b v(|\ul r+\ul u-\ul v|)
\end{split}
\eeq
Note that as $g(r)$ and $v(r)$ are rotationally invariant, so are $F_p(r)$ and $F_v(r)$. 
If $\r(\ul u)$ is given by Eq.~(\ref{rho}), one gets
\beq
F(|\ul r|) = \int d\ul u \, \frac{e^{-\frac{u^2}{4A}}}{(\sqrt{4\p A})^d} f(|\ul r+\ul u|)
\eeq
where $f(r) \in \{g(r), \, g(r)\log g(r), \, g(r) \b v(r)\}$.
For Hard Spheres $F_v \equiv 0$.

\section{Small cage expansion}
\label{sec:smallcageexp}

The strategy of~\cite{MP99} was to expand the HNC free energy in a power series
of the cage radius $A$, assuming that the latter is small close to the glass
transition. The expansion is carried out easily if the pair potential $v(r)$ and
the pair correlation $g(r)$ are analytic functions of $r$. However this is not
the case for Hard Spheres, as $g(r)$ vanishes for $r < D$ and has a discontinuity
in $r=D$, so the formulae of~\cite{MP99} for the power
series expansion of $\Psi$ cannot be applied to our system. 
In this section, we will
work out the expansion in the case where the pair correlation $g(r)$ has 
discontinuities. 

It is crucial to realize, that independently from any approximation, in the limit $A \to 0$, the
partition function becomes (neglecting a trivial factor) the partition function of a single atom
at an effective temperature given by $\b_{eff}=\b m$. In the case of hard spheres, where there is no
dependence on the temperature, the change in temperature is irrelevant.

In \cite{MP99} it was shown that the first term of the expansion is proportional to $A$ if $g(r)$ is
differentiable.  As we will see in the following, in the case of hard spheres, the presence of a
jump in $g(r)$ produces terms $O(\sqrt{A})$ in the expansion.  In this paper we will focus on these
terms neglecting all the contributions of higher order in $\sqrt{A}$.  This means that we can
neglect all the contributions coming from the regions where $g(r)$ is differentiable and concentrate
only on what happens around $r=D$.

We will focus first on the $g (\log g-1)$ term in Eq.~(\ref{HNCfree}). The contribution
we want to estimate comes from the discontinuity of $g(r)$ in
$r=D$. Thus to compute this correction the form of $g(r)$ away from the singularity is
irrelevant and we will use the simplest possible form of $g(r)$.

\subsection{Expansion of $F_0(r)$}

First we will discuss the expansion of $F_0(r)$ in $d=1$. The simplest possible form of $g(r)$ is
\beq
g(r)=\theta(r-D) [1+ (y-1) e^{-\m (r-D)}] \ ;
\eeq
the amplitude of the jump of $g(r)$ in $r=D$ is given by $y$.
Remember that in our notation $\ul r \in \RRR$ and $r=|\ul r| \in \RRR^+$.
As the functions $F_0$ and $g$ are even in $\ul r$, we can write
\beq
\label{termine1}
\int_{-\io}^\io d\ul r [F_0(\ul r)^m - g(\ul r)^m] = 2 \int_0^\io dr [F_0(r)^m - g(r)^m] \ .
\eeq
Defining
\beq
\begin{split}
&\erf(t) \equiv \frac{2}{\sqrt{\p}} \int_0^t dx \, e^{-x^2} \ , \\
&\Th(t) = \frac{1}{2} [ 1 + \erf(t) ] \ ,
\end{split}
\eeq
these functions play the role of ``smoothed'' sign and $\th$-function
respectively; note also that the function $\Th(t)$ goes to $0$ as $e^{-t^2}$ for
$t\rightarrow -\io$. Then
\beq\begin{split}
&\int_{-\io}^\io du \, \frac{e^{-\frac{u^2}{4A}}}{\sqrt{4\p A}} \,
\theta(r+u-D) = \\ &\frac{1}{2} 
\left[ 1 + \text{erf}\left(\frac{r-D}{\sqrt{4A}}\right) \right] 
\equiv \Th\left(\frac{r-D}{\sqrt{4A}}\right) \ ,
\end{split}\eeq
and
\beq
\label{F0}
\begin{split}
F_0(r)&=\Th\left(\frac{r-D}{\sqrt{4A}}\right)
+ \Th\left(-\frac{r+D}{\sqrt{4A}}\right) \\
 &+ (y-1) e^{A \m^2}
\Big\{ e^{-\m(r-D)} \Th\left(\frac{r-D-2A\m}{\sqrt{4A}}\right)\\
& + e^{\m(r+D)}\Th\left(-\frac{r+D+2A\m}{\sqrt{4A}}\right)
\Big\} \ .
\end{split}
\eeq
As $r\geq 0$ we can neglect the terms proportional to $\Th\left(-\frac{r+D}{\sqrt{4A}}\right)$
in Eq.~(\ref{F0}), that give a contribution of order $\exp(-D^2/A)$ for $A \rightarrow 0$.
Defining the reduced variable $t= (r-D)/\sqrt{4A}$:
\beq
\begin{split}
g(t) &= \theta(t) [1+(y-1) e^{-\m 2 \sqrt{A} t}] \ , \\
F_0(t)&=\Th(t) + (y-1)  e^{-\m 2 \sqrt{A} t} \, e^{A \m^2}\Th(t+\m \sqrt{A}) \ ,
\end{split}
\eeq
and Eq.~(\ref{termine1}) becomes
\beq\begin{split}
&\int_0^\io dr [F_0(r)^m - g(r)^m] =\\
&2\sqrt{A} \int_{-\frac{D}{\sqrt{4A}}}^\io dt [ F_0(t)^m - g(t)^m ] \equiv 2 \sqrt{A} Q(A) \ .
\end{split}\eeq
If the function $Q(A)$ has a finite limit $Q(0)$ for $A\rightarrow 0$ we will have 
$Q(A) = Q(0) + o(1)$ and the leading correction to the free energy is 
$O(\sqrt{A} Q(0))$.
The limit for $A \rightarrow 0$ of $Q(A)$ is formally given by
\beq
Q(0) = y^m \int_{-\io}^\io dt \, [ \Th(t)^m - \th(t)^m ] \equiv y^m Q_m
\eeq
where $y^m \equiv Y$ is the jump of $G(r)\equiv g(r)^m$ in $r=D$ and 
$Q_m \equiv \int_{-\io}^\io dt \, [ \Th(t)^m - \th(t)^m ]$.
It is easy to show that $Q_m$ is a finite and smooth function of $m$ for
$m \neq 0$, that
\beq
\begin{split}
&Q_m = (1-m) Q_0 + O[(m-1)^2] \ , \\
&Q_0 = -\int_{-\io}^\io dt \, \Th(t) \log \Th(t) \sim 0.638 \ ,
\end{split}
\eeq
and that $Q_m$ diverges as $Q_m \sim \sqrt{\p/4m}$ for $m \rightarrow 0$.
Finally we get, recalling that $G(r)=[g(r)]^m$,
\beq
\label{Gterm}
\frac{1}{2}\int d\ul r \, F_0(r)^m = \frac{1}{2}\int d\ul r \, G(r) + 2 \sqrt{A} Y Q_m \ .
\eeq
In dimension $d>1$ we have, recalling that $F_0(r)$ and $G(r)$ are both rotationally invariant,
\beq
\label{DDDD}
\int d\ul r \, [ F_0(r)^m - G(r)^m ] = \O_d \int_0^\io dr \, r^{d-1} \, [ F_0(r)^m - G(r)^m ] \ ,
\eeq
where $\O_d$ is the solid angle in $d$ dimension, $\O_d=2\pi^{d/2}/\G(d/2)$.
The function $F_0(r)$ can be written as
\beq
F_0(r) = \int d\ul u \, \frac{e^{-\frac{u^2}{4A}}}{(\sqrt{4\p A})^d} g(|r \widehat i +\ul u|) \ ,
\eeq 
where $\widehat i$ is the unit vector e.g. of the first direction in $\RRR^d$. For small $\sqrt{A}$, 
the $u$ are small too. The function $g(|r \widehat i +\ul u|)$ is differentiable along the directions
orthogonal to $\widehat i$. Expanding in series of $u_\m$, $\m \neq 1$, at fixed $u_1$, we see that
the integration over these variables gives a contribution $O(A)$, so we finally get:
\beq
\label{5D}
F_0(r) = \int_{-\io}^\io du_1 \, \frac{e^{-\frac{u_1^2}{4A}}}{\sqrt{4\p A}} g(r + u_1) + O(A) \ ,
\eeq 
as in the one dimensional case. The function $F_0(r)^m - G(r)^m$ is large only for
$r - D \sim \sqrt{A}$ so at the lowest order we can replace $r^{d-1}$ with $D^{d-1}$
in Eq.~(\ref{DDDD}). We get
\beq
\int d\ul r \, [ F_0(r)^m - G(r)^m ] = \O_d D^{d-1} \int_0^\io dr \, [ F_0(r)^m - G(r)^m ] \ .
\eeq
The last integral, with $F_0(r)$ given by Eq.~(\ref{5D}) is the same as in $d=1$, so we obtain
\beq
\label{GtermD}
\frac{1}{2} \int d\ul r \, F_0(r)^m = \frac{1}{2} \int d\ul r \, G(r) + 
\sqrt{A} Y \Si_d(D) Q_m \ ,
\eeq
where $\Si_d(D)$ is the surface of a $d$-dimensional sphere of radius $D$,
$\Si_d(D) = \O_d D^{d-1}$.
This result can be formally written as
\beq\begin{split}
\label{corrQ}
F_0(r)^m &\sim G(r) + 2\sqrt{A} Y Q_m \d(|r|-D) \\
&\equiv G(r) + Q_0(r)
\end{split}\eeq
as the correction comes only from the region close to the singularity of 
$g(r)$, $r-D \sim \sqrt{A}$.

\subsection{$G\log G$-term}

Let us now estimate the correction coming from the term 
$\int dr \, m F_0(r)^{m-1} F_1(r)$.
Using the same argument as in the previous subsection, we will
restrict to $d=1$.
Note first that $F_0(r)$, for $|r-D| \sim \sqrt{A}$, has the form
\beq
\label{F0sing}
F_0(r) = y \, \Th\left(\frac{r-D}{\sqrt{4A}}\right) + o(\sqrt{A}) \ ,
\eeq
where $y$ is the jump of the function $g(r)$ in $r=D$. Similarly,
$F_1(r)$ will have the form
\beq 
F_1(r) =
\begin{cases} 
g(r) \log g(r) + O(A) \ , \hskip33pt  |r-D| \gg \sqrt{A} \ , \\
y\log y \, \Th\left(\frac{r-D}{\sqrt{4A}}\right) + o(\sqrt{A}) \ , \hskip10pt
|r-D| \sim \sqrt{A} \ .
\end{cases}
\eeq
The integral 
\beq
\int_0^\io dr [ m F_0(r)^{m-1} F_1(r) - m g(r)^m \log g(r) ]
\eeq
has then two contributions: the first comes from the region $|r-D| \gg \sqrt{A}$ and
is of order $A$ as if the function $g(r)$ were continuous. The other comes from the
region $|r-D| \sim \sqrt{A}$ and is of order $\sqrt{A}$ as in the previous case.
To estimate the latter we can use again the reduced variable $t$ and approximate
$F_1(t) \sim y \log y \, \Th(t)$, $F_0(t) \sim y \, \Th(t)$.
Then we get
\beq\begin{split}
\int_0^\io dr& [ m F_0(r)^{m-1} F_1(r) - m g(r)^m \log g(r) ] =\\
&Y \log Y \, 2\sqrt{A} Q_m + o(\sqrt{A}) \ ,
\end{split}\eeq
in $d=1$ and finally, in any dimension $d$,
\beq
\label{GlogGtermD}
\begin{split}
&\frac{1}{2}\int d\ul r \, m F_0(r)^{m-1} F_1(r) =\\
&\frac{1}{2}\int d\ul r \, G(r) \log G(r) + \sqrt{A} Y \log Y \, \Si_d(D) Q_m \ .
\end{split}\eeq

\subsection{Interaction term}

Substituting Eq.~(\ref{rrho}) in the last term of the HNC free energy one obtains
\beq
\begin{split}
\Tr[h\r]^n &=\wh\r^n \int d\ul X_1 \cdots d\ul X_n \int du_1 \cdots du_n \times\\
&\times \r(u_1) \cdots \r(u_n)
h(\ul X_1-\ul X_2,u_1-u_2) \\& \cdots h(\ul X_n-\ul X_1,u_n-u_1) \ ,
\end{split}\eeq
where we used the notations $h(X,u) = \prod_{a=1}^m g(X+u_a) - 1$ and 
$\r(u)=\prod_{a=1}^m \r(\ul u_a)$ with
$\r(\ul u)$ given by Eq.~(\ref{rho}).

The correction $O(\sqrt{A})$ to this integral comes from the regions where 
$|X_i-X_{i+1}| = D + O(\sqrt{A})$ for some $i=1,\cdots,n$. In these regions the functions
$h$ such that their arguments are not close to the singularity can be expanded in a power series
in $u$, the correction being $O(A)$~\cite{MP99}. Thus we can write,
defining $H(r)=G(r)-1$:
\beq
\begin{split}
\wh\r^{-n} &\Tr[h\r]^n = 
\int d\ul X_1 \cdots d\ul X_n H(\ul X_1-\ul X_2) \cdots H(\ul X_n-\ul X_1) +\\
&n \int d\ul X_1 \cdots d\ul X_n \int du_1 du_2 \, \r(u_1) \r(u_2) \times\\
&\times\big[ h(\ul X_1-\ul X_2,u_1-u_2) - H(\ul X_1-\ul X_2) \big] \times\\
&\times H(\ul X_2-\ul X_3) \cdots H(\ul X_n-\ul X_1)= \\
&\int d\ul X_1 \cdots d\ul X_n H(\ul X_1-\ul X_2) \cdots H(\ul X_n-\ul X_1)\\
&+ n \int d\ul X_1 \cdots d\ul X_n Q_0(\ul X_1-\ul X_2)\times\\
&\times H(\ul X_2-\ul X_3) \cdots H(\ul X_n-\ul X_1) \ ,
\end{split}
\eeq
where in the last step we used Eq.~(\ref{corrQ}):
\beq\begin{split}
\int du_1 du_2 \, &\r(u_1) \r(u_2)
\big[ h(r,u_1-u_2) - H(r) \big] =\\& F_0(r)^m - G(r) = Q_0(r) \ .
\end{split}\eeq
Collecting all the terms with different $n$ we get
\beq
\begin{split}
&\frac{1}{2} \sum_{n\geq 3} \frac{(-1)^n}{n} \Tr[h\r]^n \sim
\frac{1}{2} \sum_{n\geq 3} \frac{(-1)^n}{n} \wh \r^n \Tr H^n + \\
&+ \frac{\wh\r^3}{2} \int d\ul X_1 d\ul X_2 d\ul X_3 Q_0(\ul X_1-\ul X_2) H(\ul X_2-\ul X_3) 
\times\\ &\times
\sum_{n \geq 3} (-1)^n \wh\r^{n-3} \int d\ul X_4 \cdots d\ul X_n H(\ul X_3-\ul X_4) \times\\
&\times \cdots H(\ul X_n-\ul X_1) = \\
&= \frac{1}{2} \sum_{n\geq 3} \frac{(-1)^n}{n} \wh \r^n \Tr H^n
- \frac{\wh\r^3}{2} \int d\ul X_1 d\ul X_2 d\ul X_3 \times\\&\times Q_0(\ul X_1-\ul X_2) H(\ul X_2-\ul X_3) C(\ul X_3-\ul X_1) \ .
\end{split}
\eeq
Substituting the expression of $Q_0(r)$ and recalling that from the definition
of $C(\ul X)$ one has 
$\wh \r \int d\ul Z H(\ul X-\ul Z) C(\ul Z-\ul Y) = H(\ul X-\ul Y) - C(\ul X-\ul Y)$, we get
\beq
\label{inttermD}
\begin{split}
&\frac{1}{2} \sum_{n\geq 3} \frac{(-1)^n}{n} \Tr[h\r]^n
=\frac{1}{2} \sum_{n\geq 3} \frac{(-1)^n}{n} \wh \r^n \Tr H^n -\\
&-N \wh\r Q_m \sqrt{A} y \Si_d(D) [ H(D)-C(D)] \ .
\end{split}\eeq
This result is correct in any dimension $d$.

\section{First order free energy}
\label{sec:freeenergy}

Substituting Eq.s~(\ref{GtermD}),~(\ref{GlogGtermD}) and (\ref{inttermD}) in Eq.~(\ref{HNCfree2})
one obtains the following expression for
the HNC free energy at first order in $\sqrt{A}$:
\beq
\begin{split}
\b& F = \frac{\b \Psi}{N} = \b F_0(A) + \b F_{eq}[G(r)] + \b \D F[A,G(r)] \ , \\
\b& F_{eq} = \frac{\wh\r}{2} \int d^d r \, \{ G(r) \log G(r) - G(r) + 1 \} \\
&+ \frac{1}{2\wh\r} \int \frac{d^d k}{(2\p)^d} 
\left[ - \log[1+\wh H(k)] + \wh H(k) - \frac{1}{2} \wh H(k)^2 \right] \\
&+ \log \wh\r -1 \ , \\
\b &F_0 = \frac{d}{2} (1-m) \log (2\p A) + \frac{d}{2} (1-m) 
- \frac{d}{2} \log m \ , \\
\b &\D F = \wh\r Q_m \sqrt{A} \Si_d(D) \, G(D) \times\\
&\times \big[ \log G(D) - 1 - H(D) + C(D) ) \big] \ ,
\end{split}
\eeq
where $Q_m = Q_0 (1-m) + o((m-1)^2)$, $Q_0 \sim 0.638$ and the Fourier transform has been defined as
\beq
\label{Fdef}
\wh H(k) = \wh\r \int dr \, e^{i kr} H(r) \ .
\eeq

At the first order in $\sqrt{A}$ we only need to know the function $G(r)$ determined by the
optimization of the free energy at the zeroth order in $\sqrt{A}$, i.e. the usual free energy
$F_{eq}[G(r)]$: it satisfies the HNC equation $\log G(r) = H(r) - C(r)$.  Substituting this
relation in $\b \D F$ one simply obtains $\b \D F = -\wh\r Q_m \sqrt{A} \Si_d(D) \, G(D)$.

The derivative w.r.t. $A$ leads to the following expression for the cage radius:
\beq
\label{Amd}
\sqrt{A^*} = \frac{1-m}{Q_m} \frac{d}{\wh\r \Si_d(D) G(D)}
\eeq
which in $d=3$ becomes (let us define again $Y=G(D)$):
\beq
\label{Am}
\frac{\sqrt{A^*}}{D} = \frac{1-m}{Q_m} \frac{1}{8 \, \f \, Y(\f)}
\eeq
where $\f = \frac{\pi D^3 \wh \r}{6}$ is the {\it packing fraction}. Substituting this result
in $\b \D F$ one has $\b \D F(A^*) =d(m-1)$.

\begin{figure} 
\centering 
\includegraphics[width=.45\textwidth,angle=0]{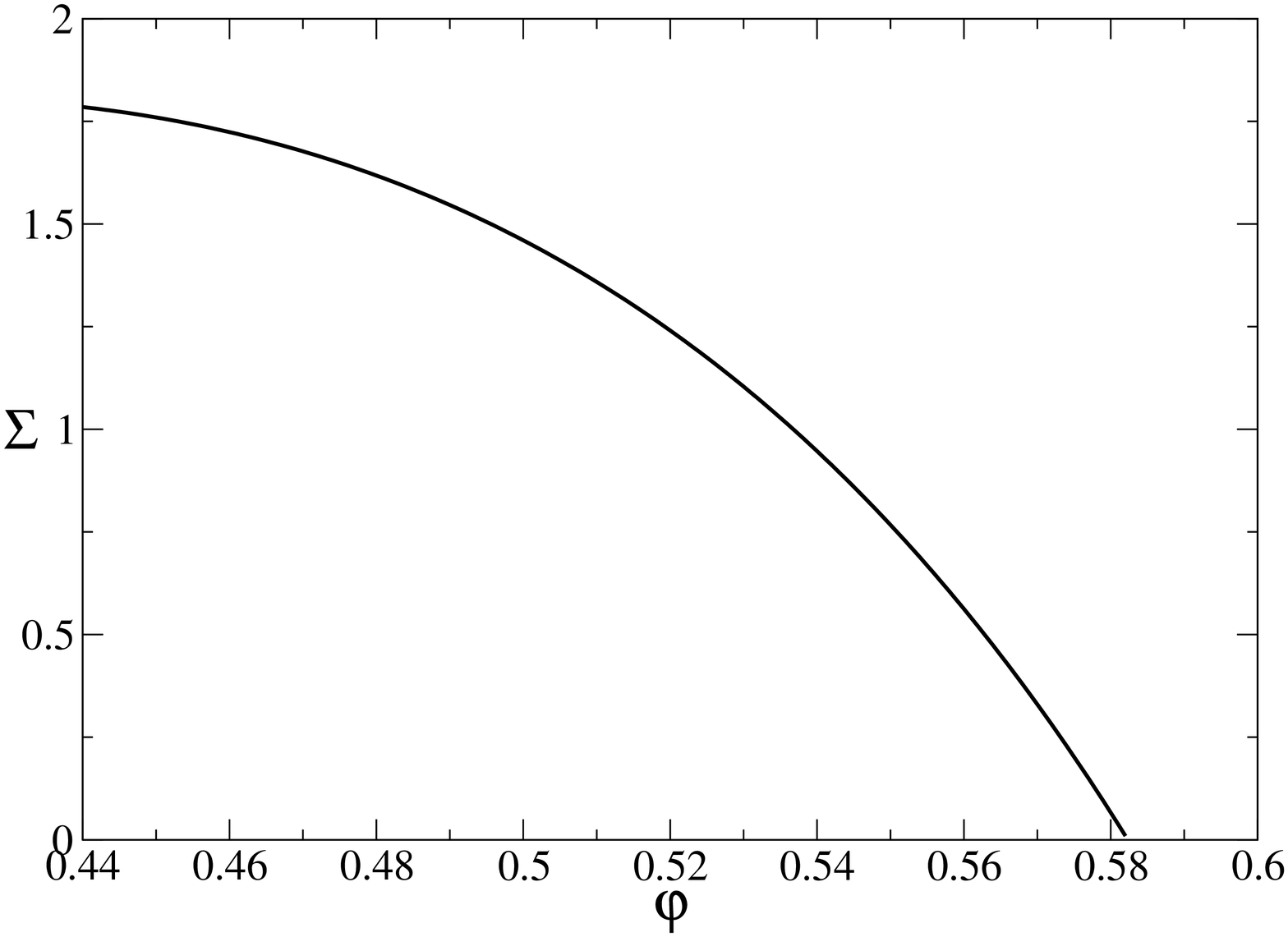}
\caption{The equilibrium complexity $\Si(\f)$ as a function of the packing fraction.}
\label{fig:Sieq} 
\end{figure} 

Finally, the expression for the replicated free energy in $d=3$ is
\beq
\label{Phim}
\begin{split}
\b \Phi(m,\f) &= \b F_{eq}(\f) + \frac{3}{2} (1-m) \log [ 2\p A^*(m)] \\
&+ \frac{3}{2}(m-1) - \frac{3}{2} \log m
\end{split}\eeq
Note that for Hard Spheres one has $\b F_{eq}(\f) = -S(\f)$, $S$ being the total entropy 
of the liquid. We get then
\beq
\label{Sm}
\begin{split}
\b f^*(&m,\f) = \frac{\partial \b \Phi}{\partial m} = 
-\frac{3}{2} \log [2\p A^*(m)] \\
&+ \frac{3}{2} (1-m) \frac{d \log A^*(m)}{dm} + \frac{3}{2} \frac{m-1}{m} \ , \\
\Si(m&,\f) = m \b f^* - \b\Phi = S(\f) - \frac{3}{2} \log [ 2\p A^*(m)] \\
&+\frac{3m}{2} (1-m) \frac{d \log A^*(m)}{dm}+ \frac{3}{2} \log m
\end{split}
\eeq
For small enough density the system is in the liquid phase and  $m$ is equal to 1 at the saddle point.
For $m=1$ we have:
\beq
\label{Sm1}
\begin{split}
&\frac{\sqrt{A^*(1)}}{D} = \frac{1}{8 Q_0 \, \f \, Y(\f)} \\
&S_{vib}(\f) \equiv -\b f^*(1,\f) = \frac{3}{2} \log [2\p A^*(1)] \\
&\Si(\f) = S(\f) - S_{vib}(\f) 
\end{split}
\eeq
This allows for a computation of $\Si(\f)$ once $S(\f)$ and $Y(\f)$ are known.
Note that $1+4\f Y(\f)=\b P/\r=-\f \frac{\partial S}{\partial \f}$, so a model for $S(\f)$
(or $Y(\f)$) is enough to determine all the quantities of interest.

\section{Results from the HNC free energy}
\label{sec:results}

We computed numerically $S(\f)$ and $Y(\f)$ solving the classical HNC equation for the 
Hard Sphere liquid up to $\f=0.65$.
This allows to compute $\b\Phi(\f,m)$ and gives access to all the thermodynamic quantities
using Eq.s~(\ref{Sm}) and (\ref{Sm1}).
In this section we discuss the results of this computation. We will set the sphere
diameter $D=1$ in the following.

\subsection{Equilibrium complexity}

The equilibrium complexity $\Si(\f)$ is given by Eq.~(\ref{Sm1}). It is reported
in Fig.~\ref{fig:Sieq}. We get a complexity $\Si \sim 1$ as found in previous
calculations in Lennard-Jones systems~\cite{MP99,CMPV99,MP99b,MP00}, as well as in
the numerical simulations~\cite{CMPV99,SKT99}. The complexity vanishes at $\f_K = 0.582$, that
is the ideal glass transition density --or Kauzmann density-- predicted by
the HNC equations.

\subsection{Phase diagram in the $(\f,m)$ plane}

\begin{figure} 
\centering 
\includegraphics[width=.45\textwidth,angle=0]{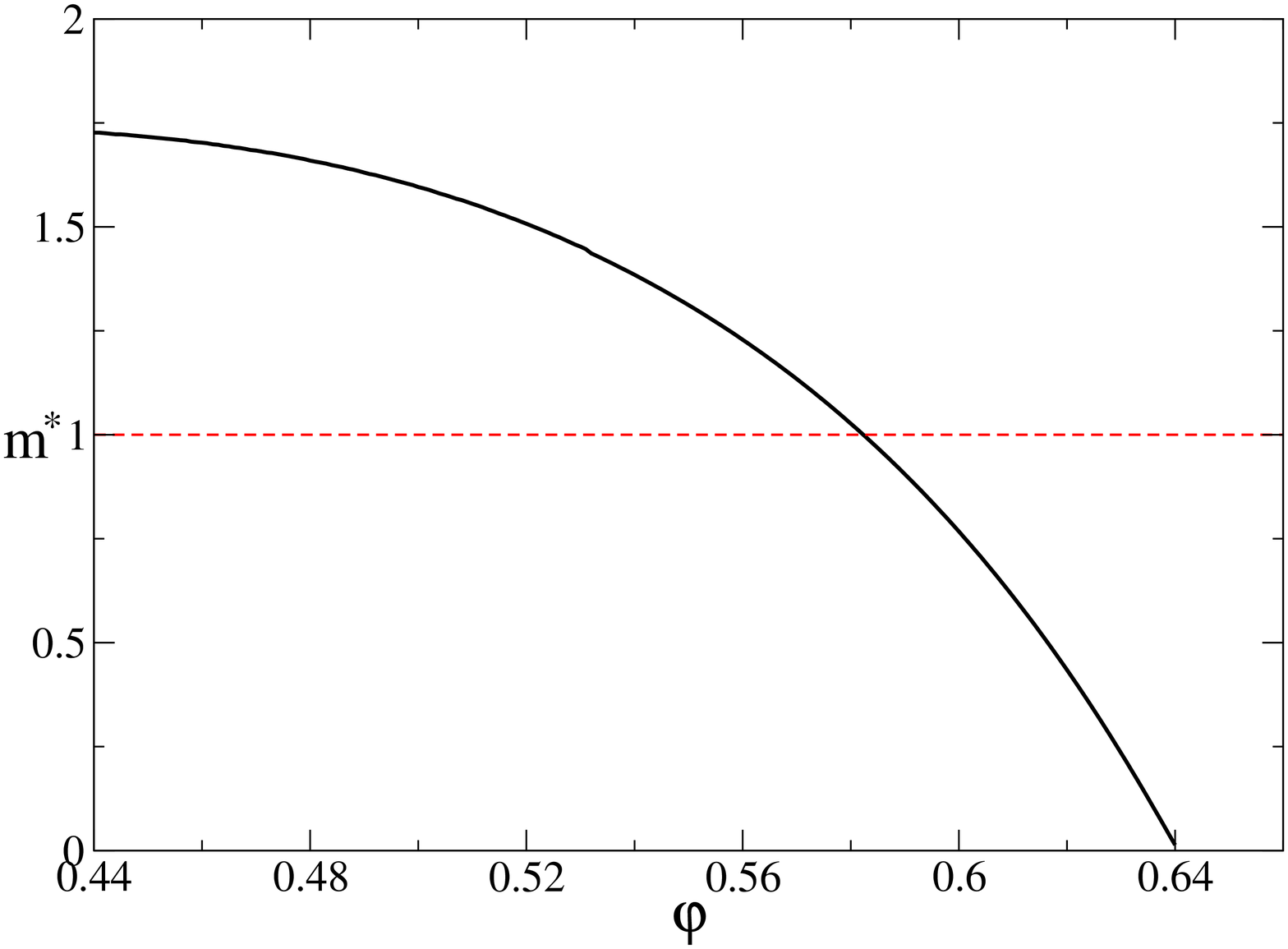}
\caption{Phase diagram of the molecular liquid. For $m<m^*$ (full line)
the system is in the liquid phase, for $m>m^*$ it is in the glass phase.
}
\label{fig:mstar} 
\end{figure} 
We now compute the thermodynamic properties of the glassy 
phase for $\f > \f_K$.
As discussed above, it exists a value of $m$, $m^*(\f)$, such that
for $m < m^*(\f)$ the system is in the liquid phase. It is the
solution of $\Si(m,\f)=0$, where $\Si(m,\f)$ is given by
Eq.~(\ref{Sm}).
In Fig.~\ref{fig:mstar} we report $m^*$ as a function of $\f$.
Clearly, $m^*=1$ at $\f=\f_K$ and $m^* < 1$ for $\f > \f_K$.
$m^*$ vanishes linearly at $\f_c=0.640$. As we will see in the
following, above this value of $\f$ the glassy state does not
exist anymore.

\subsection{Thermodynamic properties of the glass}

\begin{figure} 
\centering 
\includegraphics[width=.5\textwidth,angle=0]{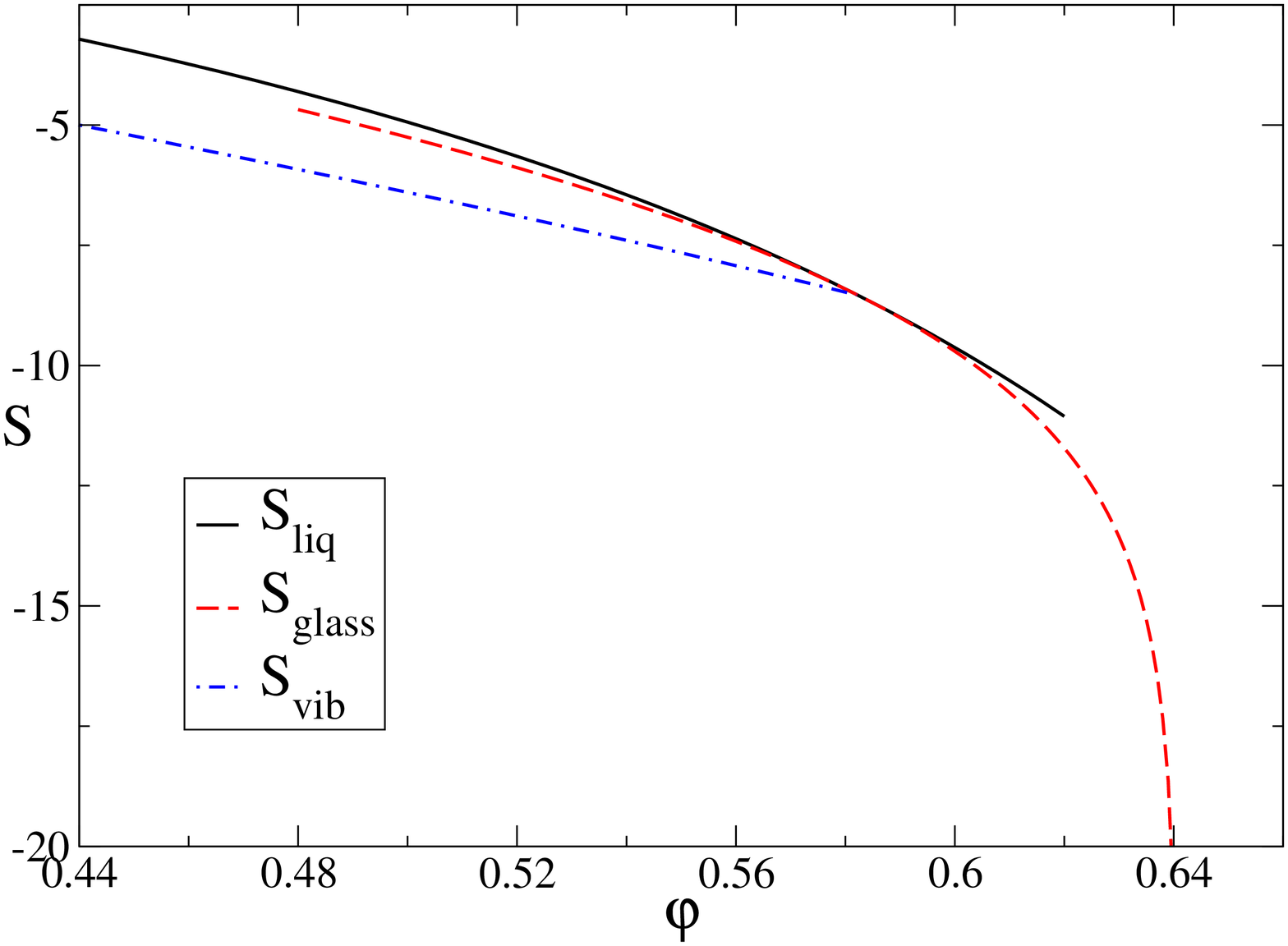}
\caption{Entropy of the liquid (full line) and of the glass
(dashed line). The two curves intersect at $\f_K=0.582$ where
they are tangent and consequently the pressure is continuous
at the glass transition. The entropy of the glass goes to 
$-\io$ at $\f=\f_c=0.640$, so the glassy phase does not exist above
$\f_c$. The dot--dashed line is the entropy of the equilibrium
states of the liquid, $S_{vib}(\f)=S(\f)-\Si(\f)$.
}
\label{fig:entropy} 
\end{figure}
The knowledge of the function $m^*(\f)$ allows to compute
the entropy of the glass. Indeed, the free energy does not
depend on $m$ in the whole glassy phase, and it is continuous
along the line $m=m^*(\f)$, so we can
compute the entropy of the glass simply as
\beq
\label{Sglass}
S_{glass}(\f)=-\b F_{glass}(\f)=-\frac{\b \Phi(m^*(\f),\f)}{m^*(\f)}
\eeq
This relation is true for $m^* < 1$. Below $\f_K$ one has $m^* > 1$
and the liquid phase is the stable one. Eq.~(\ref{Sglass}) for $m^*>1$ 
gives the entropy of the lowest states in the free energy landscape 
(see below) and can be regarded as the analytic
continuation of the glass entropy below $\f_K$. The reader should notice that
the glass phase for $m^*>1$ does not have a simple physical meaning and the interesting part of
the curves for the glass is in the region $\f>\f_K$.

In Fig.~\ref{fig:entropy} we report the entropies of the liquid and the
glass as functions of the packing fraction. The glass phase becomes
stable above $\f_K = 0.582$; note that the entropy of the glass is
{\it smaller} than the entropy of the liquid, {\it i.e.} its free
energy is {\it bigger} than the free energy of the liquid.
The same happens also in Lennard-Jones systems and in mean-field
spin glass systems. However the physical relevant parts of the curves are the liquid one for
$\f<\f_K$ and the glassy one for $\f>\f_K$.

\begin{figure} 
\centering 
\includegraphics[width=.5\textwidth,angle=0]{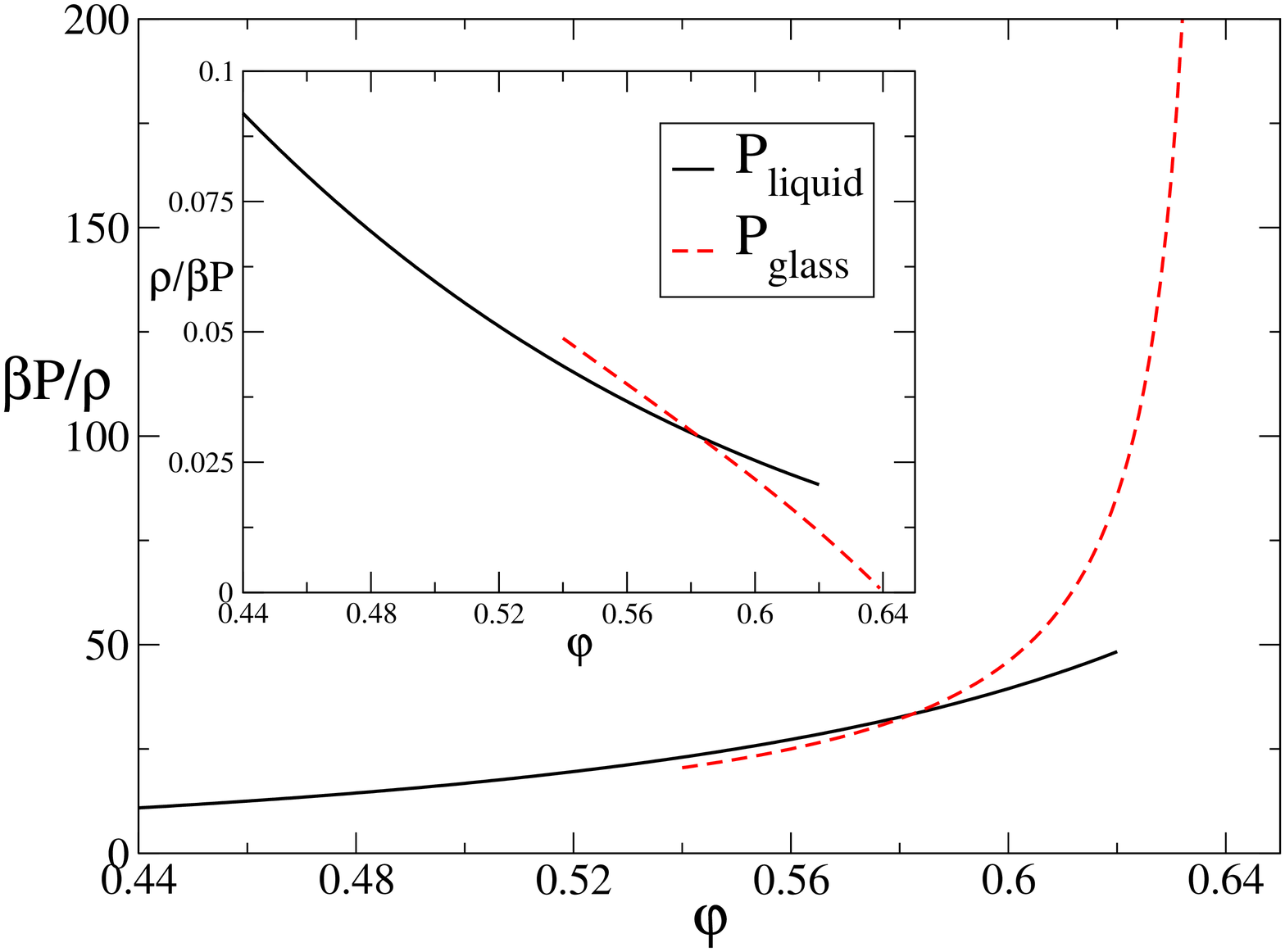}
\caption{Reduced pressure $\b P/\r$ of the liquid and the glass as functions
of the packing fraction. The pressure is continuous at $\f_K$.
In the inset, the inverse reduced pressure is plotted; in the glass
phase it is proportional to $\f_c-\f$.
}
\label{fig:pressure} 
\end{figure}

The reduced pressure,
\beq
\frac{\b P}{\r} = -\f\frac{\partial S}{\partial \f} \ ,
\eeq
is reported in Fig.~\ref{fig:pressure}.
It is continuous at $\f_K$ and  the glass transition
is a second order transition from the thermodynamical point of view.
Note that the pressure in the glass phase is well described by a
power law and it has a simple pole at $\f_c$:
\beq
\frac{\b P_{glass}}{\r} \propto \frac{1}{\f_c-\f} \ ,
\eeq
as one can see from the inset of Fig.~\ref{fig:pressure} where the inverse
reduced pressure is plotted as a function of $\f$.

For $\f \rightarrow \f_c$ the pressure of the glass diverges and its
compressibility 
$\chi=\frac{1}{\f}\frac{\partial \f}{\partial P}$ vanishes and consequently
 $\f_c$ is the maximum density allowed for a disordered
state, {\it i.e.} it can be identified as the 
{\it random close packing density}.
The value $\f_c=0.640$ is in very good agreement with the values
reported in the literature.
Note that the compressibility jumps downward on increasing $\f$ across
$\f_K$, {\it i.e.} the compressibility of the glass is smaller than the
compressibility of the liquid.

\subsection{Cage radius}

\begin{figure} 
\centering 
\includegraphics[width=.5\textwidth,angle=0]{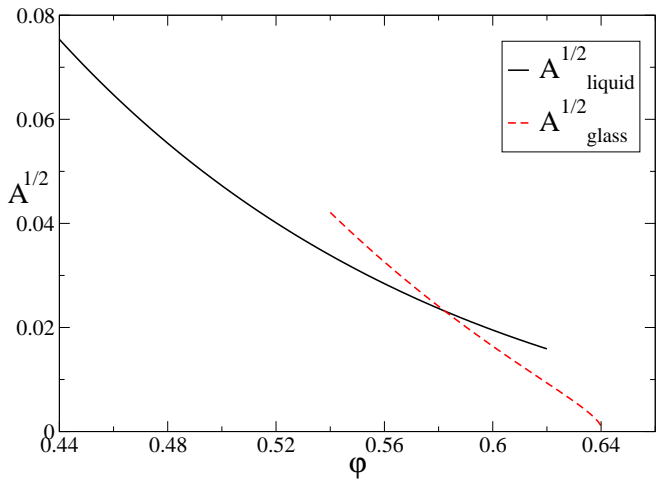}
\caption{
Cage radius $\sqrt{A}$ (in units of $D$) 
in the liquid and in the glass phase as function of $\f$.
}
\label{fig:A} 
\end{figure}

The cage radius is given as a function of $m$ in Eq.~(\ref{Am}). 
In Fig.~\ref{fig:A} we report the cage radius in the liquid phase,
$\sqrt{A^*(1)}$, see Eq.~(\ref{Sm1}), and the cage radius in the
glass phase, defined as $\sqrt{A^*(m^*)}$. As $Q_m \sim \sqrt{\p/4m}$
for $m\sim 0$, the cage radius vanishes as $\sqrt{m^*}$ for $m^* \sim 0$,
{\it i.e.} it is proportional to $\sqrt{\f_c-\f}$.
The vanishing of the cage radius for $\f \rightarrow \f_c$ means that
at $\f_c$ each sphere is in contact with its neighbors, that is consistent
with our interpretation of $\f_c$ as the random close packing density.

\subsection{Complexity of the metastable states}

\begin{figure} 
\centering 
\includegraphics[width=.47\textwidth,angle=0]{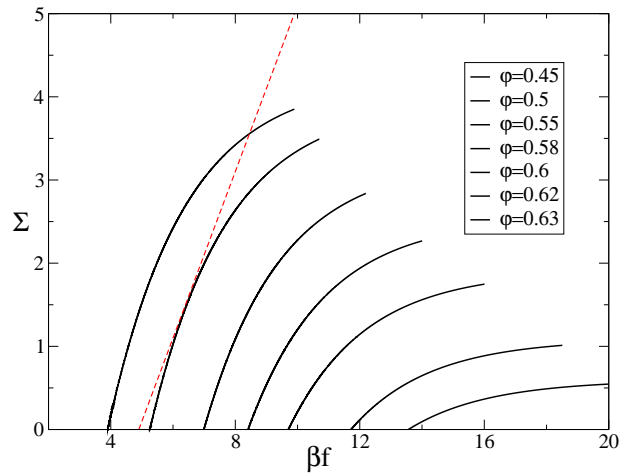}
\caption{
Complexity of the metastable states as a function of their free energy
$\b f$ for different values of $\f$.
From left to right, $\f=0.45, 0.5, 0.55, 0.58, 0.6, 0.62, 0.63$.
The curves are truncated arbitrarily at high $\b f$.
The dashed line has slope~$1$.
}
\label{fig:Sif} 
\end{figure}

From the parametric plot of $\b f^*(m,\f)$ and $\Si(m,\f)$ given in
Eq.~(\ref{Sm}) by varying $m$, one can reconstruct the function 
$\Si(\b f)$ for each value of the packing fraction. 
This function is reported in
Fig.~\ref{fig:Sif} for some values of $\f$ below and above $\f_K$.
The function $\Si(\b f)$ vanishes at a certain value $\b f_{min}$,
that is given by Eq.~(\ref{Sglass}). The saddle-point equation that
determines the free energy of the equilibrium states is, from
Eq.~(\ref{Zm1}),
\beq
\label{saddleeq}
\frac{d \Si(\b f)}{d \b f} = 1 \ .
\eeq
From Fig.~\ref{fig:Sif} we see that this equation has a solution
$f^* > f_{min}$ for $\f < \f_K = 0.582$. 
Above $\f_K$ Eq.~(\ref{saddleeq}) does not have a solution so the
saddle point is simply $f^* = f_{min}$ and the systems goes in
the glass state. In this sense, the free energy $f_{min}$ of the
lowest states below $\f_K$ can be regarded as the analytic 
continuation of the free energy of the glass, see 
Fig.~\ref{fig:entropy}.
The curves $\Si(\b f)$ in Fig.~\ref{fig:Sif} have been truncated 
arbitrarily at high $\b f$. We have not done consistency checks to 
investigate where the higher free energy states become unstable
(\ie, to compute $f_{max}$).

\section{Correlation functions}
\label{sec:correlations}

We will now turn to the study of the pair distribution function 
$\tilde g(r)$ in the glass state. In principle a full computation would require the 
evaluation of the corrections proportional to $\sqrt{A}$ in the correlation 
functions of a molecule. 
However we neglect these terms, that we believe are small, and we consider again our
simple {\it ansatz} (\ref{rrho}), (\ref{gprod}) for the correlation function of the
molecules, in which the information on the shape of the molecule is only encoded
in the function $\r(x)$;
these corrections should be physically more relevant and interesting.

As we will see in the following, the correlation function of the spheres in the glass is very
similar to the one in the liquid but develops an additional strong peak (that becomes a
$\d$-function at $\f_c$) around $r=D$.  The integral of the latter peak is related to the average
coordination number of the random close packings.

\subsection{Expression of $\tilde g(r)$ in the glass phase}

We assumed the following form for the pair distribution function
of the molecular liquid, see Eq.s~(\ref{rrho}) and (\ref{gprod}):
\beq
\label{r2}
\begin{split}
&\r_2(x,y)=\r(x) g(x,y) \r(y) =\\
&\wh\r^2 \int dX dY \prod_{a=1}^m \r(x_a-X) g(|x_a-y_a|) \r(y_a-Y) \ .
\end{split}
\eeq
The pair correlation $\tilde g(r)$ of a single replica is obtained 
integrating over the coordinates of all the replicas but one:
\beq
\tilde g(|x_1-y_1|) = \wh \r^{-2} \int d\ul x_2\cdots d\ul x_m 
d\ul y_2 \cdots d\ul y_m \r_2(x,y) \ .
\eeq
Using Eq.~(\ref{r2}) we get, with some simple changes of variable:
\beq
\tilde g(r) = g(r) \int d\ul u d\ul v \r(\ul u) \r(\ul v) F_0(|\ul r+\ul u-\ul v|)^{m-1} \ ,
\eeq
where $F_0(r)$ is defined in Eq.~(\ref{Fp}). The HNC free energy is
optimized by $g(r)=G(r)^{1/m}$, where $G(r)$ is the HNC pair correlation.
Thus we get the following expression for the pair correlation of a single replica:
\beq
\label{gvetro}
\begin{split}
&\tilde g(r) = G(r)^{\frac{1}{m}} \int d\ul u \frac{e^{-\frac{u^2}{4A}}}{(\sqrt{4\p A})^d} 
F_0(|\ul r+\ul u|)^{m-1} \ , \\
&F_0(r) =  \int d\ul u \frac{e^{-\frac{u^2}{4A}}}{(\sqrt{4\p A})^d} G(|\ul r+\ul u|)^{\frac{1}{m}} \ .
\end{split}
\eeq
For $m=1$, {\it i.e.} in the liquid phase, this function is trivially equal
to $G(r)$. This is not the case in the glass phase where $m<1$.

\subsection{Small cage expansion of the correlation function}

We will now expand Eq.~(\ref{gvetro}) for small $A$. Note first that, if $r \neq D$,
the function $g(r+u)$ can be expanded in powers of $u$, and the first correction
to $\tilde g(r)$ is of order $A$. Then, as before, we will concentrate on what happens
around $r=D$.
As already discussed in section~\ref{sec:smallcageexp}, around $r=D$ we have,
as in Eq.~(\ref{F0sing}), $G(r) \sim Y \th(r-D)$ and
\beq
F_0(r) \sim Y^{\frac{1}{m}} \Th\left(\frac{r-D}{\sqrt{4A}}\right) \ ,
\eeq
and Eq.~(\ref{gvetro}) becomes
\beq
\tilde g(r) = Y \th(r-D) \int d\ul u \frac{e^{-\frac{u^2}{4A}}}{(\sqrt{4\p A})^d} 
\Th\left(\frac{|\ul r+\ul u|-D}{\sqrt{4A}}\right)^{m-1} \ .
\eeq
Applying the same argument we used in section~\ref{sec:smallcageexp} when studying 
the function $F_0(r)$ in dimension $d > 1$, we can show that the integration over
the coordinates $u_\m$, $\m \neq 1$, gives a contribution $O(A)$. Then we can
rewrite, in any dimension $d$:
\beq
\label{gvetro2}
\begin{split}
&\tilde g(r) \sim Y \th(r-D) \int_{-\io}^\io du \frac{e^{-\frac{u^2}{4A}}}{\sqrt{4\p A}}
\Th\left(\frac{r+u-D}{\sqrt{4A}}\right)^{m-1} \\
&= G(r) \left\{ 1 + \int_{-\io}^\io \frac{dt}{\sqrt{\p}} 
e^{-\left(\frac{r-D}{\sqrt{4A}}-t\right)^2}
 \left[ \Th(t)^{m-1} - 1 \right]\right\} \ ,
\end{split}
\eeq
defining the reduced variable $t=\frac{r+u-D}{\sqrt{4A}}$.
The second term in the latter expression is a contribution localized
around $r = D$.

\subsection{Number of contacts}

To compute the average number of contacts, let us recall that the 
average number of particles in a shell $[r,r+dr]$, if there
is a particle in the origin, is given by
\beq
dn(r) = \O_d r^{d-1} \wh\r \, \tilde g(r) dr \ .
\eeq
Thus the number of contacts can be obtained from the correlation function $\tilde g(r)$.
While the full computation of the correlation function is rather involved, 
here we limit ourselves to consider the second term in Eq.~(\ref{gvetro2}),
which is proportional to a Gaussian with variance $O(\sqrt{A})$ that becomes 
a $\d(|r|-D)$-function in the limit $A \rightarrow 0$.

The value of the number of spheres in contact with the sphere in the
origin is given by
\beq
z = \O_d \wh\r \int_D^{D+O(\sqrt{A})} dr \, r^{d-1} \tilde g(r) \ .
\eeq
The first term in Eq.~(\ref{gvetro2}) gives a contribution $O(\sqrt{A})$
that can be neglected. If we use $r \sim D$ and $G(r) \sim Y$ at the
leading order in $\sqrt{A}$ we obtain,
defining the variable $\epsilon = \frac{r-D}{\sqrt{4A}}$,
\beq
\begin{split}
&z = \O_d D^{d-1} \wh\r Y  \times \\ &\times \sqrt{4A} \int_0^\io d\epsilon 
\int_{-\io}^\io \frac{dt}{\sqrt{\p}} 
e^{-(\epsilon-t)^2} \left[ \Th(t)^{m-1} - 1 \right] \ .
\end{split}
\eeq
Recalling that
\beq
\frac{1}{\sqrt{\p}}\int_0^\io d\epsilon \, e^{-(\epsilon-t)^2} = \Th(t) \ ,
\eeq
we get, observing that $\int_{-\io}^\io dt \, \big[ \Th(t) - \th(t) \big] = 0$,
and using Eq.~(\ref{Amd}),
\beq
\label{nD}
\begin{split}
z &= \Si_d(D) \wh\r Y  \sqrt{4A}
\int_{-\io}^\io dt \, \Th(t) \left[ \Th(t)^{m-1} - 1 \right] \\
&=  \Si_d(D) \wh\r Y  \sqrt{4A} Q_m = 2 d (1-m) \ .
\end{split}
\eeq
This is the expression of the average number of contacts at the leading
order in $\sqrt{A}$, to be computed at $m=m^*$ in the glass phase.
At $\f=\f_c$, where $m^*=0$, each sphere has on average $2d$ contacts. 
This is exactly what is found in numerical simulations; the condition $z\geq 2d$ is
required for the mechanical stability of the packings as can be understood
by mean of a very simple argument~\cite{Al98}.

Note that this result is independent on the particular expression we chose for
$S(\f)$, $Y(\f)$ and $G(r)$, {\it i.e.} it might hold beyond the 
choice of HNC equations for the molecular liquid provided that
the expression~(\ref{Amd}) for the cage radius is correct.

%%%%%%%%%%%%%%%%%%%%%%%%%%%%%%%%%%%%%%%%%%%%%%%%%%%%%%%%%%%%%%%%%%%%%%%%
%%%%%%%%%%          DISCUSSION         %%%%%%%%%%%%%%%%%%%%%%%%%%%%%%%%%
%%%%%%%%%%%%%%%%%%%%%%%%%%%%%%%%%%%%%%%%%%%%%%%%%%%%%%%%%%%%%%%%%%%%%%%%

\section{Discussion}
\label{sec:discussion}

We will now compare our results with related ones that appeared in
the literature. The main obstacle for a quantitative comparison is that
the HNC equations are known to yield a not very good description of the 
Hard Sphere liquid at high density~\cite{Hansen}; typically one would obtain the right curves if
one shifts the value of $\f$ of a quantity of order 0.03.  Therefore, we should limit 
ourselves to a {\it qualitative} comparison of the results coming from 
the HNC equations with the results of numerical simulations.
However, note that, although the 
expressions~(\ref{Am}), (\ref{Phim}) for the replicated free energy
have been derived starting from the expression (\ref{HNCfree}) for
the HNC free energy, the final result depends only on the equilibrium
entropy of the liquid $S(\f)$. It is interesting then, for the purpose of 
comparing our results with experiments and numerical simulations, 
to consider a more accurate model for $S(\f)$ in the liquid phase. 
We repeated the calculations of section~\ref{sec:results}
substituting the Carnahan--Starling (CS) entropy~\cite{Hansen} 
\beq
\begin{split}
&S_{CS}(\f)=-\log\left(\frac{6\f}{\p e}\right)-\frac{4\f-3\f^2}{(1-\f)^2} \ , \\
&Y_{CS}(\f)=\frac{1-\frac{1}{2}\f}{(1-\f)^3} \ .
\end{split}
\eeq
instead of the HNC entropy in Eq.s~(\ref{Phim}), (\ref{Am}).
All the results of section~\ref{sec:results} are qualitatively reproduced
using the CS entropy, but the latter gives results in better agreement
with the numerical data. However, this procedure is not completely
consistent from a theoretical point of view: one should always keep
in mind that our aim here is not to present a quantitative theory,
but only to show that the replica approach yields a reasonable 
qualitative scenario for the glass transition in Hard Sphere systems.

\subsection{Complexity of the liquid and Kauzmann density}

\begin{figure} 
\centering 
\includegraphics[width=.45\textwidth,angle=0]{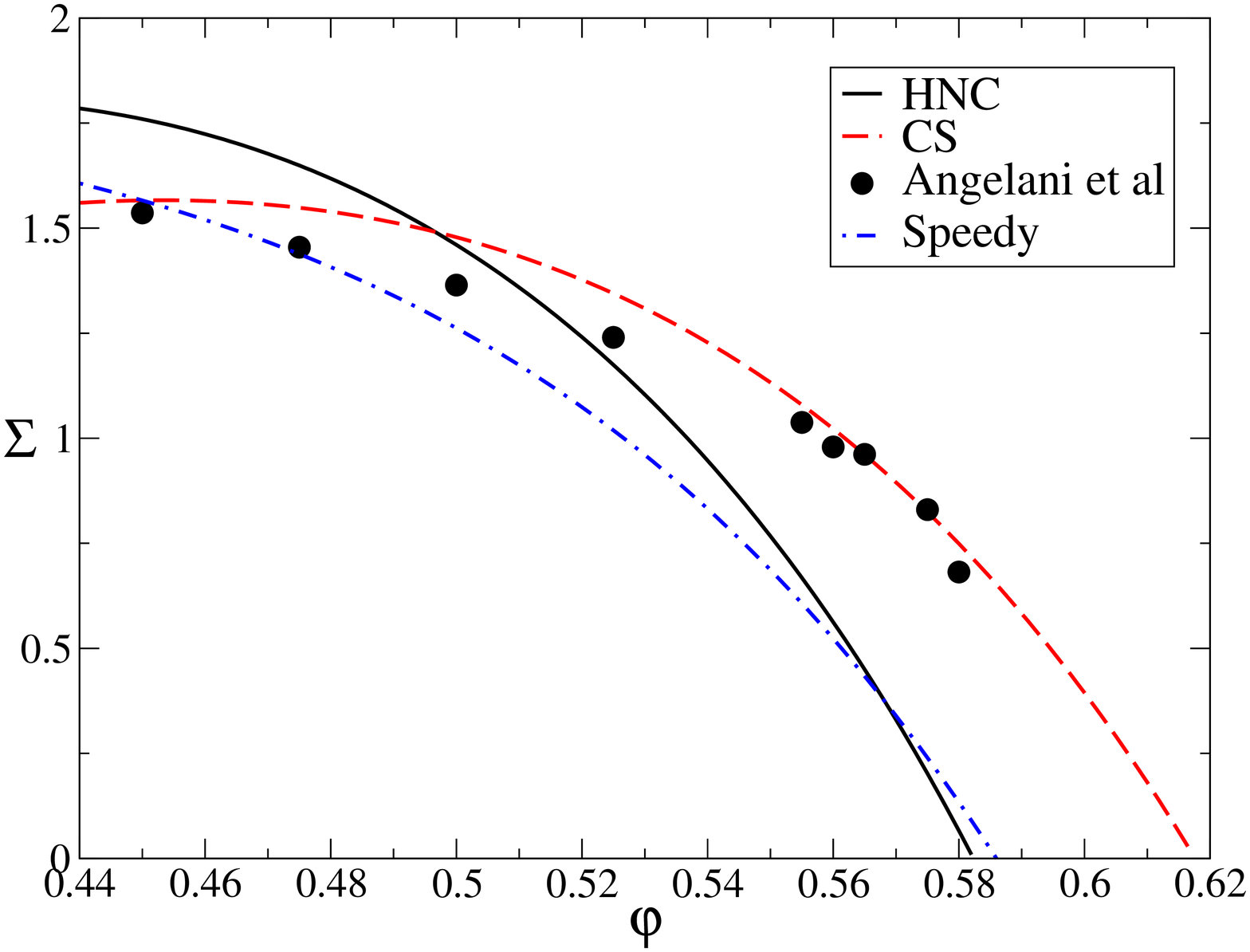}
\caption{Equilibrium complexity $\Si(\f)$ as a function of the packing fraction.
The full line is from the HNC equation of state (see Fig.~\ref{fig:Sieq}), 
the dashed line is from the
Carnahan--Starling equation of state. The black dots are numerical data
of Angelani {\it et al.}~\cite{Luca05}
(the data shown here correspond to $\a_0^{(1)}$ in Ref.~\cite{Luca05}), 
the dot--dashed line
is extrapolated from the numerical data reported by Speedy~\cite{Sp98}.
}
\label{fig:Sicomp} 
\end{figure} 
In Fig.~\ref{fig:Sicomp} we report the equilibrium complexity $\Si(\f)$
obtained substituting the HNC and the CS expression for $S(\f)$ and $Y(\f)$ 
in Eq.~(\ref{Sm1}). The results are compared with recent numerical results
of Angelani {\it et al.}~\cite{Luca05} obtained on a $50:50$ binary mixture 
of spheres (to avoid crystallization) with diameter ratio equal to $1.2$:
the vibrational entropy was estimated using the procedure described 
in~\cite{CMPV99,AFST04} and the complexity was computed as $S(\f)-S_{vib}(\f)$.
A quantitative comparison is difficult here because in the case of a mixture
there can be corrections related to the mixing entropy, $S_{mix} \sim \log 2$.
Nevertheless the data are in good agreement with our results. A detailed
comparison would require the extension of our computation to binary mixtures
following~\cite{CMPV99}.

Another numerical estimate of $\Si(\f)$ was previously reported by 
Speedy~\cite{Sp98}, who
rationalized his numerical data assuming a Gaussian distribution of
states and a particular form for the vibrational entropy inside a
state. The free parameters were then fitted from the liquid equation
of state. The curve obtained by Speedy also agrees with our results.

Both the HNC and the CS estimates of the Kauzmann density ($\f_K=0.582$ and
$\f_K=0.617$ respectively) fall, as it should be,
between the Mode--Coupling dynamical transition that is
$\f_{MCT}\sim 0.56$~\cite{GS91,vMU93}, and the Random Close Packing
density that is estimated in the range $\f=0.64\div 0.67$, see e.g.~\cite{Be83}.

A computation of $\Si(\f)$ based on very similar ideas was presented
in~\cite{CFP98}, where a very similar estimate of $\f_K \sim 0.62$ was
obtained. However in \cite{CFP98} the complexity was found to be
$\Si \sim 0.01$, {\it i.e.} two orders of magnitude smaller than
the one obtained from the numerical simulations.
This negative result is probably due to some technical problem
in the assumptions of~\cite{CFP98}.

\subsection{Equation of state of the glass}

\begin{figure} 
\centering 
\includegraphics[width=.45\textwidth,angle=0]{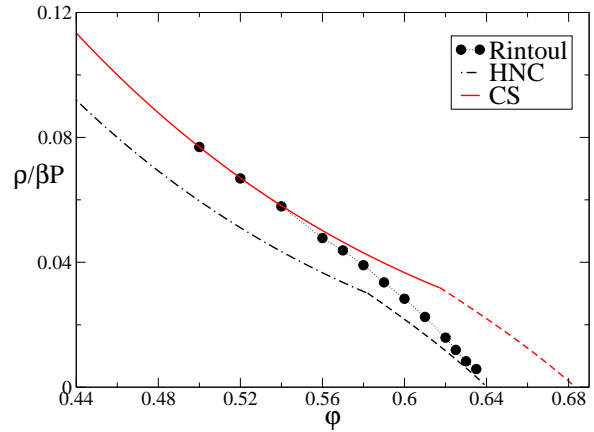}
\caption{Inverse reduced pressure $\frac{\r}{\b P}$ of the Hard Sphere 
liquid as a function of $\f$. The black dots are from the simulation
of Rintoul and Torquato \cite{RT96}. The full line is obtained from
the CS equation of state while the dot--dashed line is from the HNC
equation of state. The dashed parts of the two curves correspond to
the (ideal) glass phase. Note that all the curves are quasi--linear 
functions of $\f$ in the glass phase.
}
\label{fig:Pcomp} 
\end{figure}
In Fig.~\ref{fig:Pcomp} we report as black dots the
numerical data for the pressure of the Hard Sphere liquid 
at high $\f$ obtained by Rintoul and Torquato~\cite{RT96}. 
The data were obtained extrapolating at long times the relaxation
of the pressure as a function of time after an increase of density
starting from an equilibrated configuration at lower density.
We also report the curves of the pressure as a function of the
density obtained from the HNC and CS equations, both in the liquid
and in the glass state.

The agreement of the HNC curve with the data is not very good even
in the liquid phase, due to the modest accuracy of the HNC equation of state.
However, the qualitative behavior of our curve is in good agreement
with the numerical data, and in particular the quasi--linear behavior of
the inverse reduced pressure in the glass phase found in~\cite{RT96,Sp98},
$\frac{\r}{\b P} \propto \f_c-\f$,
is reproduced by the HNC curve. The HNC pressure of the glass diverges
at $\f_c=0.640$ as discussed in section~\ref{sec:results}; the latter
is the HNC estimate of the random close packing density.

The CS curve describes well the pressure in the liquid phase~\cite{Hansen}.
Comparing the curve with the data of Rintoul and Torquato, we see that the glass
transition happens in the numerical simulation at a density $\f_g\sim 0.56$
smaller than the one predicted by the CS curve, $\f_K=0.617$~\cite{nota1},
and very close to the Mode--Coupling transition density, $\f_{MCT}\sim 0.56$.
This is not surprising, since the relaxation time grows
fast on approaching the ideal glass transition; at some point it becomes 
larger than the experimental time scale and the liquid falls 
out of equilibrium becoming a {\it real} glass. It is likely that the data of 
Ref.~\cite{RT96} describe the pressure of a real {\it nonequilibrium} glass, while 
our computation gives the pressure of the ideal {\it equilibrium} glass,
that cannot be reached experimentally in finite time.

\subsection{Random close packing}

Both the HNC and CS equations predict the existence of a {\it random close packing} density $\f_c$
where the pressure and the value of the radial distribution function $\tilde g(r)$ in $r=D$ diverge.
The HNC estimate is $\f_c=0.640$, in the range of the values ($\f_c=0.64\div 0.67$) reported in the
literature. The CS estimate is $\f_c=0.683$ and it is also a value consistent with numerical
simulations.

The reader should notice that the theoretical value for $\f_c$ is related to the 
{\it ideal} random close packing;
however the states corresponding to this value of $\f_c$ can be reached by local algorithms, like
most of the algorithms that were used in the literature, in a time that should diverge exponentially
with the volume. Some caution should be taken in using the data obtained by numerical simulations.
The question of which is the value of the density that can be obtained in large, but finite amount of time
per particle is very interesting and more relevant from a practical point of view: however we plan
to study it at a later time.

Note that the computation of the mean coordination number $z$ 
of section~\ref{sec:correlations}, that gives $z=6$ at $\f=\f_c$ in $d=3$,
is {\it independent} of the particular form we choose for $S(\f)$,
and thus is valid for both the HNC and CS equations of state.
The value $z=6$ has been reported in many studies
\cite{Be72,Ma74,Po79,Al98,SEGHL02}.

\section{Conclusions}

We successfully applied the replica method of~\cite{Mo95,MP99} to
the study of the ideal glass transition of Hard Spheres, and 
in general of potentials such that the pair distribution
function $g(r)$ shows discontinuities, starting from
the replicated HNC free energy and expanding it at first
order in the cage radius $\sqrt{A}$.

This result allowed us to compute from first principles the 
configurational entropy of the liquid as well as 
the thermodynamic properties of the glass up to the
random close packing density. Our computation is based
on the HNC equation of state, that is known to yield a
poor quantitative description of the liquid state at high
density. Nevertheless, we found that the qualitative scenario
for the ideal glass transition that emerges from the replicated
HNC free energy is very reasonable. In particular, we found
a complexity $\Si \sim 1$, a Kauzmann density $\f_K =0.582$, and
a random close packing density $\f_c=0.64$. All these results
compare well with numerical simulations.

Using, on a phenomenological ground, the Carnahan--Starling
equation of state instead of the HNC equation of state as
input for our calculations, we could also compare 
our results with the high--density pressure data 
of Rintoul and Torquato showing that they are indeed compatible
with the observation of a real glass transition.

Moreover, we found that the mean 
coordination number in the amorphous packed states is $z=2d$
irrespective of the equation of state we use for the liquid,
in very good agreement with the result of numerical 
simulations and with theoretical 
arguments~\cite{Be72,Ma74,SEGHL02,Al98}.

It is worth to note that our results do not {\it prove} the 
existence of a glass transition for the Hard Sphere liquid,
as they derive from a particular approximation for the molecular
liquid free energy (the HNC approximation), and, in general, other 
approximation such as the Percus--Yevick are possible~\cite{Hansen}.

\acknowledgments

We are grateful to L.~Angelani, G.~Foffi and F.~Sciortino for providing their
data prior to publication and for their comments on this work.
F.Z. wish also to thank E.~Zaccarelli for the code for solving the HNC
equations and for many interesting discussions.


\begin{thebibliography}{99}
\bibitem{nota1} Note that the authors of \cite{RT96} interpreted their
data as showing no evidence for a glass transition, the pressure being
a differentiable function of $\f$. However, as recognized in
\cite{RLSB98}, their data are better described by a broken curve
showing a glass transition around $\f_g=0.56$.
\end{thebibliography}
\end{document}